\documentclass[10pt,a4paper]{article}
\usepackage[utf8]{inputenc}
\usepackage{amsmath}
\usepackage{amsfonts}
\usepackage{amssymb}
\usepackage{graphicx}
\usepackage{siunitx}
\usepackage[square,sort,comma,numbers]{natbib}
\usepackage[toc,page]{appendix}
\usepackage[nottoc,numbib]{tocbibind}
\numberwithin{equation}{section}
\title{\vspace{-15mm}\fontsize{12pt}{10pt}\selectfont\textbf{Volumetric velocimetry for small seeding tracers in large volumes}\author{Yisheng Zhang  \and
        Simon L. Ribergård \and
        Haim Abitan \and
        Clara M. Velte}  } 
\begin{document}

\maketitle

\begin{abstract}
 This paper presents the volumetric velocity measurement method of small seeding tracer with diameter  $5\,\mathrm{\mu m}\sim 100\,\mathrm{\mu m}$ for volume $\geq 500\,\mathrm{cm}^3$. The size of seeding tracer is between helium-filled soap bobbles (HFSB)  and di-ethyl-hexyl-sebacic acid ester(DEHS) droplets. The targeted measurement volume dimension equal to the volume of HFSB seeding, which will give a higher resolution of turbulence study. The relations between particle size, imaging and light intensity are  formulated. The estimation of the imaging result are computed for the setup design. Finally, the methodology is demonstrated for turbulence velocity measurements in the jet flow, in which the velocities of averaged diameter $15 \,\mathrm{\mu m}$ air filled soap bubbles are measured in the volume of $\geq 2000\,\mathrm{cm}^3$ and $\geq 9000\,\mathrm{cm}^3$.

\end{abstract}

\section{Introduction}
\begin{figure}
  \includegraphics[width=\textwidth]{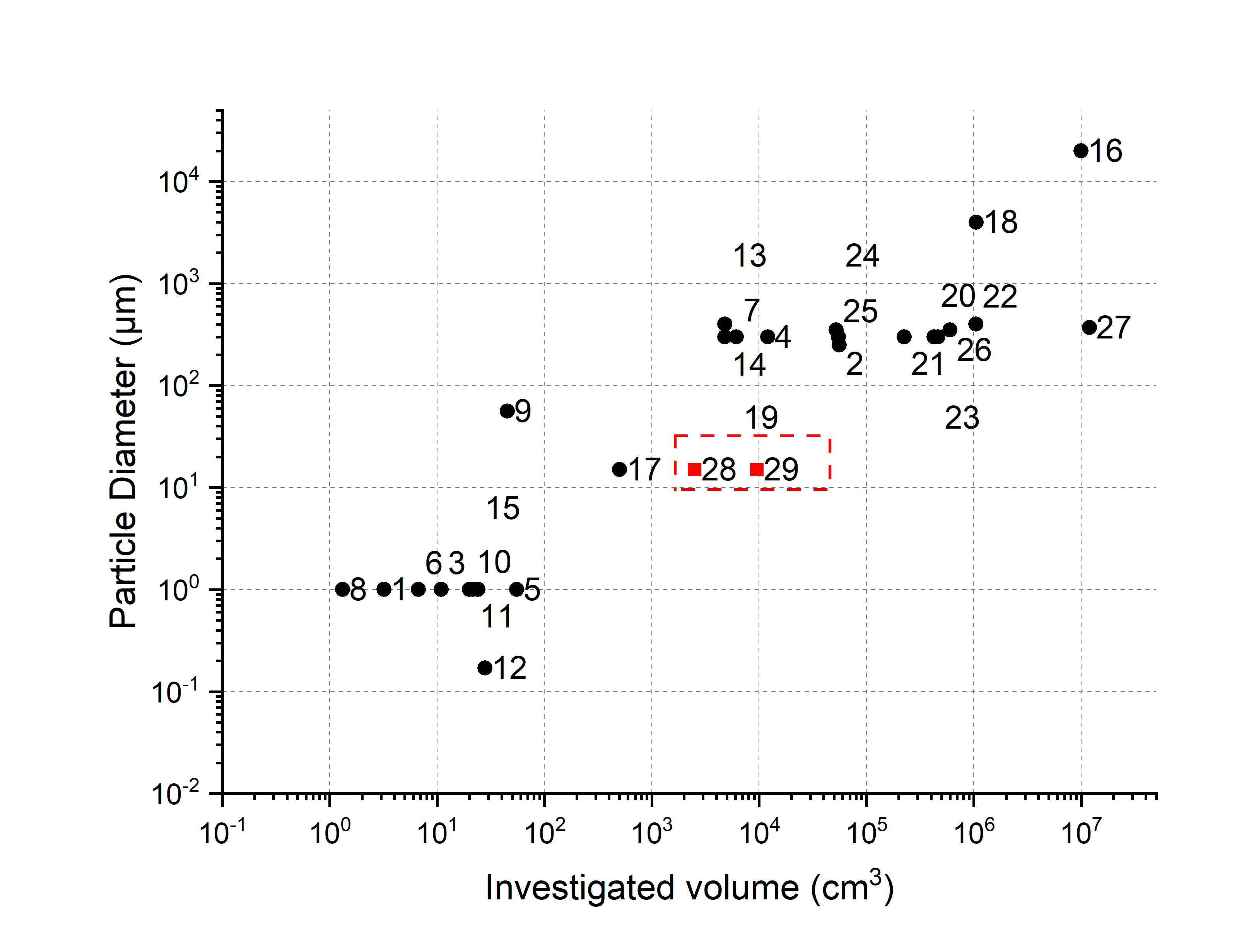}
\caption{Compilation of volumetric velocimetry investigations in air flows: the particle diameter and the imaged volume. The labeled points refers to the researches in the references: 1, Atkinson et al.\cite{Atkinson2011}; 2, Kühn et al.\cite{Kuehn2011}; 3, Cafiero et al.\cite{Cafiero2015}; 4, Caridi et al.\cite{Caridi2016}; 5, Boushaki et al.\cite{Boushaki2017}; 6, Pröbsting et al.\cite{Probsting2013}; 7, Schneiders et al.\cite{Schneiders2016A}; 8, Schneiders et al.\cite{Schneiders2016B}; 9, Schneiders et al.\cite{Schneiders2016B}; 10, Elsinga et al.\cite{Elsinga2006}; 11, Schröder et al.\cite{Schroeder2007}; 12, Humble et al.\cite{Humble2009}; 13, Terra et al.\cite{Terra2017}; 14, Scarano et al.\cite{Scarano2015}; 15, Michaelis et al.\cite{Michaelis2012}; 16, Hou et al.\cite{Hou2021}; 17, Barros et al.\cite{Barros2021}; 18, Bosbach et al.\cite{Bosbach2009}; 19, Schneiders et al.\cite{Schneiders2016}; 20, Huhn et al.\cite{Huhn2017}; 21, Schanz et al.\cite{Schanz2018}; 22, Bosbach et al.\cite{Bosbach2019}; 23, Schanz et al.\cite{Schanz2019}; 24, Schanz et al.\cite{Schanz2016}; 25, Wolf et al.\cite{Wolf2019}; 26, Novara et al.\cite{Novara2021}; 27, Schröder et al.\cite{Schroeder2021}. The red squares 28 and 29 are the current study and the red area is the interested study area. }
\label{fig:compilation} 
\end{figure}

It is still a problem in the research of turbulent flows for obtaining all the eddies from the largest primary size to a micron size in inertial sub-range. The main conflicts come from the tracer size, camera resolution and the intensity of light signal. There is a strong push for large volumetric velocity measurement using small tracers\cite{Barros2021}. Normally, due to light scattering and characterics length scale considerations, big tracers are used for tracing large eddies in big volumes, and smaller tracers are applied for micro-scale eddies in smaller volumes. As can be expected from Figure \ref{fig:compilation}, there is a noticeable gap between attainable volumetric domains for large ($\geq 300\,\mathrm{cm}^3$) and small ($1\,\mathrm{\mu m}\sim 20\,\mathrm{\mu m}$) tracers in the range between $10^4\,\mathrm{cm}^3 \sim 10^4\,\mathrm{cm}^3$. \\
As part of a long-term combined theoretical and experimental initiative, we are setting out to experimentally test the degree of locality of the interactions between wavenumber components \cite{Ribergard2021}. This requires measurements that cover the ‘global’ dynamics of the flow, while also capturing as wide a bandwidth of the turbulent (temporal and spatial) spectrum as possible. Tracers of diameter $\sim 15\,\mathrm{\mu m}$ with reasonable Stokes number can follow the flow accurately to the degree of spatio-temporal resolution in the experimental design while at the same time they are highly resilient and long-lived and therefore the most viable option from a practical point of view. Therefore, these smaller seeding particles are the key factor to these experiments.\\
Discetti and Coletti\cite{Discetti2018} summarized methods that have been developed to measure the full volumetric velocity (3D–3C, i.e. three-dimensional and three-component). In the past two decades, volumetric velocimetry has evolved from experiments at moderate scale of $10 \,\mathrm{cm}^3$ up to large scales of $10^4\,\mathrm{cm}^3$ by various application experiments. The initial tomo-PIV experiment done by Elsinga et al. \cite{Elsinga2006} was conducted with $1\,\mathrm{\mu m}$ droplets in a volume of $13\,\mathrm{cm}^3$. Later the larger measurement volumes were obtianed by volumetric velocimetry in air conducted by Schröder et al.\cite{Schroeder2009}, Humble et al.\cite{Humble2009}, Violato et al. \cite{Violato2011} and Ghaemi and Scarano \cite{Ghaemi2011} Schröder et al. \cite{Schroeder2011} and Atkinson et al. \cite{Atkinson2011}, which did not exceed $10^2\,\mathrm{cm}^3$ measurement volume with micron-scale seeding tracers. \\
The main factors limiting the upscale of PIV to macroscopic dimensions are the limited illumination energy from the illumination source, the scattering efficiency of the tracers, the small optical aperture (due to large depth of field), and the sensitivity and spatial resolution of the camera sensors\cite{Scarano2015}. With low repetition rate(prolonging the exposure time) and millimeter scale Helium-filled Soap Bubbles(increasing scattering efficiency), larger measurement volumes can be observed to characterize large scale flow structure \cite{Kuehn2011,Lobutova2009}.\\
For turbulence flows, the inherent instability and three dimensions requires high resolution time-resolved volumetric velocimetry. It is impossible to utilize millimeter scale tracers to measure the same scale turbulent structure, because the length scale of the structure is smaller than the tracer diameter, and the cut-off diameter of the particles deduced from cut-off frequency could be micron-scale\cite{Mei1996}. The current investigation aims at characterizing both the 'global' flow structure and millimeter scale structure with micron-scale tracers.\\
With the capabilities of low resolution camera, the only option to capture the turbulence structure is to observe small tracers in a small measurement volume. Now with the development of camera sensor techniques, the high resolution cameras in recent years provide the possibility of capturing the intermediate-size turbulence structure in a larger volume, in which the whole evolution process of the turbulent spectrum can be observed. However, there is still a problem between the visibility of seeding tracers and the following effect of intermediate-size turbulence structure. In order to visualize small structures in the turbulence flow, micron-scale tracers are needed to follow the unstable and three dimensional millimetre scale structures. And as the result of reducing the tracer dimension, the signal from seeding tracers to camera would drop in quadratic proportion.\\
For the purpose of solving this problem, this paper thus details the computation of particle image size with tracers, lens and camera parameters. Then with the desired turbulent scales, the relation among camera, light source and measurement volume is formulated to make sure that the particle images cannot travel across pixels (to avoid streaky particle images). Two examples of approximate $15\,\mathrm{\mu m}$ particles in a $> 2000 \,\mathrm{cm}^3$ volume and a $> 9000 \,\mathrm{cm}^3$ volume, that are 10 times larger than \cite{Barros2021}, are demonstrated at an acquisition frequency of $\sim 2\,\mathrm{kHz}$. The current work thus documents how to design a time-resolved volumetric velocimetry experiment in a manner that the gap of smaller seeding particles for the target turbulent microscale structure in big volume for 'global' flow is filled. \\
The paper is organized as follows. In the Section \ref{sec:method}, the methodology that related from flow characteristics, seeding, illumination and imaging was formulated. To demonstrate the computation, a jet flow example was computed in Section \ref{sec:computation}, where we use the all the formulas in Section \ref{sec:method}. The setup and experiment result corresponding to the computation in Section \ref{sec:computation} was provided in Section \ref{sec:results}. Finally, our conclusion and perspectives were given. 
\section{Methodology}
\label{sec:method}
\subsection{Flow characteristics}
The target smallest structure in the flow is characterized with length-scale $\ell$ and characteristic velocity $u(\ell)$. The timescale of these structure is \cite{Pope2000} 
\begin{equation}\label{eq:tau}
    \tau(\ell)\equiv \frac{\ell}{u(\ell)}.
\end{equation}
Based on the timescale, the interested frequency is
\begin{equation}\label{eq:fi}
    f_i=\frac{1}{\tau(\ell)}.
\end{equation}
According to the Shannon's sampling theorem, the camera minimum repetition rate $f_c$ should be more than twice of the interested frequency $f_c\geq 2f_i$. Both the spatial resolution and the camera repetition are based on the flow length-scale and timescale.
\subsection{Seeding}
There are six factors that affects the particles to follow the flow\cite{Mei1996,Adrian2011,Raffel2018}. Response time and Stokes number are related to the fidelity of tracer particles accurately following the flow\cite{Raffel2018}. Response time ($\tau_t$) represents the characteristic time required for a tracer particle to reach an equilibrium condition after a flow disturbance. The response time is commonly calculated using the following form
\begin{equation}\label{eq:taut}
    \tau_t=d^2_p \frac{\rho_p}{18\nu} 
\end{equation}
where $d_p$ is the particle diameter, $\rho_p$ is the particle density and $\nu$ is the kinematic viscosity of the ambient fluid. The Stokes number ($Sk$) is defined as the ratio of the characteristic time of a particle to a characteristic time of the interested flow ($\tau_f$),
\begin{equation}
    Sk=\frac{\tau_t}{\tau_f}
\end{equation}
As the smallest eddies are characterized the timescale $\tau(\ell)$, the flow characteristic time equals to the timescale $\tau_f=\tau(\ell)$.  In general, a Stokes number of 0.1 or less could return an acceptable flow tracing accuracy with errors below 1\% \cite{Tropea2007}.\\
In dynamic conditions like turbulent flows studied by Mei\cite{Mei1996}, the particle frequency response function $H_p$ that is defined as the sinusoidal response of particle to a sinusoidal oscillation of the flow relates to the bias between the particle velocity and flow velocity,
\begin{equation}
    \hat{v}(\omega)=H_p(\omega)\hat{u}(\omega),
\end{equation}
where $\hat{v}(\omega)$ is the particle velocity amplitude at frequency $\omega$, and $\hat{u}(\omega)$ is the flow velocity amplitude, both of which are deduced from particle oscillation function $v(t)=\hat{v}(\omega)e^{-i\omega t}$ and flow oscillation function $u(t)=\hat{u}(\omega)e^{-i\omega t}$.\\
Defining Stokes number in frequency domain $\varepsilon$ with the frequency variable $\omega$ \cite{Mei1996},
\begin{equation}
    \varepsilon=\sqrt{\frac{\omega d^2_p}{2\nu}}
    \label{eq:stokesf}
\end{equation}
the energy transfer function deduced from the particle frequency response function is
\begin{equation}
\label{eq:energy}
    \vert H_p(\omega)\vert=\vert H_p(\varepsilon)\vert=\frac{(1+\varepsilon)^2+(\varepsilon+\frac{2}{3}\varepsilon^2)^4}{(1+\varepsilon)^2+[\varepsilon+\frac{2}{3}\varepsilon^2+\frac{4}{9}(\rho-1)\varepsilon^2]^2}.
\end{equation}
where, $\rho$ is the density ratio of particles and ambient flow $\rho=\rho_p/\rho_f$, and $\rho_f$ is the density of the ambient flow.\\
In Mei's study \cite{Mei1996}, the cut-off frequencies of the particles are based the 50\% and 200\% energy response, which means that $0.5<\vert H_p(\omega)\vert^2<2$ implies very good response of the seeding particles. So the cut-off frequencies $f_{cutoff}$, the cut-off Stokes number $\varepsilon_{cutoff}$ and the cut-off particle diameter $ d_{cutoff}$ can be computed from the follow equations:
\begin{equation}
    \left\{
    \begin{array}{l}
        \varepsilon_{cutoff}=\{ \varepsilon:\vert H_p(\omega)\vert^2 \in \{ 0.5,2\}\} \\[2ex]
        \varepsilon_{cutoff}\approx [2.380^{0.93}+\bigl(\frac{0.659}{0.561-\rho}-1.175\bigr)^{0.93}]^\frac{1}{0.93},\ \rho<0.561\\[2ex]
        \varepsilon_{cutoff}\approx [\bigl(\frac{3}{2\rho^{1/2}}\bigr)^{1.05}+\bigl(\frac{0.932}{\rho-1.621}\bigr)^{1.05}]^\frac{1}{1.05}, \ \rho>1.621  \\[2ex]
        f_{cutoff}\approx \frac{\nu}{\pi}\bigl( \frac{ \varepsilon_{cutoff}}{d_p}\bigr)^2 \\[2ex]
        d_{cutoff}\approx \varepsilon_{cutoff}\sqrt{\frac{\pi  f_{cutoff}}{\nu}}
    \end{array} 
    \right.
    \label{eq:cutoff}
\end{equation}
However, in practice the computation process is the reversed version: (1)find the particle parameter like diameter $d_p$ and density $\rho_p$; (2)compute the density ratio $\rho$ and Stokes frequency domain number $\varepsilon$; (3)get the energy transfer function result $\vert H_p(\omega)\vert^2$ and decide if the seeding type is suitable for the flow; (4)compute the cut-off frequency $f_{cutoff}$ to see if it can cover the interested frequency range.\\
To emphasis the unit of the parameters in the equation, kinematic viscosity in the equation (\ref{eq:cutoff}) and (\ref{eq:stokesf}) is expressed in terms of $\mathrm{cm}^2/\mathrm{s}$, and the diameter of particles is in the term of $\mathrm{\mu m}$.
\subsection{Imaging}
The observed volume $V = L \times H \times W$ is imaged by cameras with the fixed optical magnification based on the sensor size. Define the pixel resolution $l_s$, pixel pitch $ \Delta_{pix}$, and aperture number $f_s$, the optical magnification of the imaging system can be deduced from the ratio of the focal lens to the working distance. 
\begin{equation}\label{eq:M}
    M= \frac{l_s\Delta_{pix}}{L}= \frac{d_i}{d_o}
\end{equation}
where $d_o$ is the distance from the object geometrical center to the lens center, $d_i$ the distance from the image geometrical center to the lens center. The particle image $\tau_p$ on camera sensor can be computed from:
\begin{equation}
\tau_p= \frac{\sqrt{d_{op}^2+d_s^2+d_f^2}}{\Delta_{pix}}
\end{equation} 
where $d_{op}=Md_p$ is the particle diameter; $d_s$ is diffraction limit spot diameter, which can be computed with aperture stop number $f_s$ and light wave length $\lambda$ by $d_s=2.44(1+M)f_s\lambda$, $d_f$ is the blur circle diameter.\\
The depth of focus (DOF) $\delta_z$ is
\begin{equation}
\delta_z=2d_sf_s\frac{1+M}{M^2}=4.88f^2_s\lambda\frac{(1+M)^2}{M^2}
\end{equation}
The diameter of the effective aperture $D_a$ is 
\begin{equation}
D_a=\frac{f}{f_s}
\end{equation}
And the focal length $f$ can satisfy the basic lens equation:
\begin{equation}
\frac{1}{f}=\frac{1}{d_o}+\frac{1}{d_i}
\end{equation}
\subsection{Scattered Light}
In the PIV experiment, part of the incident light is imaged by the tracers onto the camera sensors. It is the scattered light that releases the information of the tracers. The light scattering properties of a homogeneous and isotropic sphere in a plane wave, which leads to the Lorenz-Mie Theory, are well documented in the literature (van de Hulst1981\cite{Hulst1981}, Bohren and Huffmann1983\cite{Bohren1983}, Mishchenko etc.2002 \cite{Mishchenko2002}, Born and Wolf 1984\cite{Born2019}), thus only related final results are presented to obtain the light source budget. \\
The first step is to project the incident wave vector onto the scattering plane. From camera lens, the received electric field vector can be expressed as
\begin{equation}
    \mathbf{E}_r=\frac{e^{-i\mathbf{k}_\omega \mathbf{r}_p}}{\mathbf{k}_\omega \mathbf{r}_p}\mathbf{M}_\beta\mathbf{M}_S\mathbf{M}_\varphi\mathbf{E}_0
\end{equation}
where $\mathbf{E}_0$ is incident plane wave vector, $\mathbf{k}_\omega$ is the incident wave vector, and $\mathbf{r}_p$ is the scattering vector between the particle center and lens center, the angle between $\mathbf{k}_\omega$ and $\mathbf{r}_p$ is $\vartheta$, the scattering plane lies at an angle of $\varphi$, $\mathbf{M}_S$ is the scattering matrix with sum of scattering functions, $\mathbf{M}_\beta$, and $\mathbf{M}_\varphi$ is plan transform matrix (details to see Albrecht et al.\cite{Albrecht2003} pages 79-126).\\
One dimensionless parameter defined as Mie parameters is important to the scattering intensity,
\begin{equation}
    x_M=\frac{\pi d_p}{\lambda}.
\end{equation}
For particles with Mie parameter $x_M>10$, the scattered intensity increases with the second power of the particle diameter, which means geometrical optics can be applied for approximate computation. For the Lorenz-Mie region with $1\leq x_M \leq 10$, the scattered intensity exhibits strong oscillations. The received light intensity is 
\begin{equation}
    I_r(\lambda, r, \vartheta)=\frac{c\epsilon}{2}\vert\mathbf{E}_r\vert^2
\end{equation}
where $\epsilon$ is the permittivity of the medium, $c$ is the light speed.\\ 
The photons signal $I_c$ received by the cameras are the inner product of the light intensity distribution function and the camera sensor quantum efficiency function $\Phi_{qe}(\lambda)$,
\begin{equation}
    I_c=\int\int \Phi_{qe}(\lambda) I_r(\lambda, r, \vartheta) d\lambda dA.
\end{equation}
The integral of the area depends on the aperture diameter. If the distance between particle and lens is much larger than the efficient aperture $r\gg D$ and the scattering angle is in the range of geometrical optics, the scattering angle can be consider as constant. Thus the integral is only the constant intensity multiply the efficient aperture area. \\
The signal-to-noise ratio $SNR$ in this paper follows the definition of Scharnowski and Kahler\cite{Scharnowski2016}, which used standard deviations of the image intensity and the noise:
\begin{equation}
    SNR=\frac{\sigma_A}{\sigma_n},
\end{equation}
where $\sigma_n$ is the standard deviation of the noise level, and $\sigma_A$ is the standard deviation of the signal level, which can be approximated with number particle images per pixel $N_{ppp}$,
\begin{equation}
    \sigma_A=\frac{I_0}{2}\sqrt{N_{ppp}\cdot(\frac{\pi}{4}\tau_p^2-1)}
\end{equation}
The loss-of-correlation due to image noise $F_\sigma$ is defined as,
\begin{equation}
    F_\sigma=\frac{\sigma^2_A}{\sigma^2_A+\sigma^2_n}=\bigl(1+\frac{1}{SNR^2}\bigr)^{-1}.
\end{equation}
Before experiments start, the noise level $\sigma_n$ can be estimated from the camera manuals. From the theory analysis and experimental evaluation in \cite{Scharnowski2016}, the loss-of-correlation $F_\sigma$ increase strongly from 0.4 to 0.8 with linearly increment of light power. So the main focus of the light design is to pursuit the higher $F_\sigma$ below 0.8 under the limited light source budget.
\subsection{Computation process}\label{sec:computationprocess}
With all the knowledge in this section, the computation process can be summarized in the following sequence:
\begin{enumerate}
  \item Compute the camera repetition rate from the timescale of the interested flows in equation (\ref{eq:tau}) and (\ref{eq:fi}).
  \item Get the camera resolution by the ratio of smallest eddies to largest eddies, and select a camera satisfying the repetition rate and resolution.
  \item Obtain the magnification $M$ from equation (\ref{eq:M}).
  \item Set the camera position according to the distance $d_o$ computed from equation (\ref{eq:M}).
  \item Define the cut-off frequency from the reciprocal of the timescale, estimate the particle diameter scale with $\varepsilon_{cutoff}=0.1$, then compute the cut-off Stokes number $\varepsilon_{cutoff}$ and the energy transfer function $\vert H_p(\omega)\vert^2$ with different types of seeding in the same range. Set the biggest seeding type with considering the energy loss and the cost.
  \item Obtain the aperture stop number $f_s$ from depth of field of the interested volume and the diffraction limit spot diameter $d_s$ equation, and Obtain the focal length $f$ from the $d_o$ and $M$, compute  and the seeding particle image diameter $\tau_p$. 
  \item Take LEDs as the first consideration in the cases of large volumes \cite{Raffel2018}, acquire the LEDs intensity chart, and compute the Mie-parameter $x_M$.
  \item Design the light scattering angle as large as possible, the illumination distance and receiving distance as short as possible.
  \item Compute the scattering coefficient with online tools, estimate the received light intensity from 1 LED set as $I_0^{(1)}$.
  \item Set $F_\sigma=0.7$ as the basic loss-of-correlation, get $SNR=1.5$, compute the $\sigma_A$ with $\sigma_n$
  \item Compute the light intensity $I_c$, and the LEDs amount by $I_c/I_c^{(1)}$
\end{enumerate}
Following the computation process, all the components and parameters are settled from the interested volume and the interested characteristic structure scale.
\section{Experimental computation in a jet flow example}
\label{sec:computation}
In this section, we demonstrated the computation process of the visibility problem in a jet flow with approximate $15\,\mathrm{\mu m}$ diameter air-filled bubbles in a volume of $30\,\mathrm{cm}\times 30\,\mathrm{cm} \times 8\,\mathrm{cm}$. The whole estimation are computed under the condition of International Standard Atmosphere (ISA).
\subsection{Axisymmetric jet flow}
The axisymmetric jet flow is one of the best canonical flow for understanding the turbulence. It has been interrogated using both experimental and theoretical methods. Based on the experimental investigation result by Hussein, Capp and George in 1993 \cite{Hussein1994}, we model the jet velocity field in the same data model.
\begin{equation}
\frac{U_0}{U_c}=\frac{1}{B_u}\left( \frac{x}{D} - \frac{x_0}{D} \right),
\end{equation}
where the $U_0$ is the jet exit velocity, $U_c$ is the centreline velocity at position $x$, $x_0$ is the virtual origin of the jet, and $D$ is the jet nozzle diameter.\\
According to the non-dimensionalized Reynolds stresses by the square of the centreline velocity in \cite{Hussein1994}, the averaged fluctuation velocity can be computed from:
\begin{equation}
    \frac{\overline{u^2}}{U^2_c}=0.078
\end{equation}
According to measurement result in Preben Buchhave and Clara M. Velte in 2017\cite{Buchhave2017}, the length Taylor microscales are $2.2\,\mathrm{mm}$ and the temporal Taylor microscales are $1\,\mathrm{\mu s}$ at the center line at $30\,\mathrm{D}$ downstream location, which requires particle cut-off frequency higher than $1\,\mathrm{kHz}$ and camera repetition rate higher than $2\,\mathrm{kHz}$. From 13mm radial distance off the jet centerline, the temporal Taylor microscales are $2\,\mathrm{\mu s}$,  which requires particle cut-off frequency $500\,\mathrm{Hz}$.\\
\subsection{Seeding computation}
The seeding tracers that we use to verify the method are the air-filled micron-scale bubble tracers, which have mean diameters of approximately $15\,\mathrm{\mu m}$. Diogo C. Barros etc.\cite{Barros2021} investigated the characteristics of the seeding in a wind tunnel experiment. The theoretical response time computed from equation (\ref{eq:taut}) is $20\,\mathrm{\mu s}$, and the measured mean response time was $40\,\mathrm{\mu s}$.  More details of the microbubble tracer and the bubble generator can be found in \cite{Barros2021}\\
The reason we choose this type of seeding is that the tracers size is big enough to reflect light signal, and is small enough to show trajectory of Taylor microscale, and the response time is small enough to follow Taylor micro-scale. As the temporal Taylor micro-scales is $1\,\mathrm{\mu s}$ at the center line at 30D downstream location for 1cm jet at $Re=19868$\cite{Buchhave2017}, the Stocks number of the micro-bubbles was found to be 0.04, which is in the acceptable upper limit of 0.1 \cite{Tropea2007}.\\
According to Kerho and Bragg \cite{Kerho1994} and Afanasyev et al. \cite{Afanasyev2011}, the wall thickness of soap bubbles lies in the range $0.1\sim 0.3\,\mathrm{\mu m}$, the density of the bubbles is $20\sim 70 \,\mathrm{kg}\cdot\mathrm{m}^{-3}$\cite{Barros2021}. Taking the visible bubble diameter from $10\sim 30\,\mathrm{\mu m}$, which have the frequency domain Stoke number at 0.26 and 0.47, the cut-off frequencies varies from $752.2\,\mathrm{Hz}$ for large high density bubbles to $6.77\,\mathrm{kHz}$ for small low density bubbles. And the mean cut-off frequency is $2.21\,\mathrm{kHz}$ for $15\,\mathrm{\mu m}$ particles with $0.2\mu m$ wall thickness. So in the radial range of $13\,\mathrm{mm}$ at 30D downstream, only minor parts of the tracers (2.5\% computed from statistics in \cite{Barros2021}), of which the diameter is larger than $30\,\mathrm{\mu m}$ and the wall thickness is thicker than $0.3\,\mathrm{\mu m}$, cannot represent the Taylor micro-scale in the $10\,\mathrm{mm}$ jet. Out the radial range of $13\,\mathrm{mm}$ at $30\,\mathrm{D}$ downstream, all the tracers follows the Taylor micro-scale eddies well. For the average diameter particles, the energy transfer function in equation (\ref{eq:energy}) equals to 0.7088, which is in the efficient response range $[0.5, 2]$.\\
\subsection{Imaging computation}
Considering the sensor size $27.6\,\mathrm{mm}\times 26.3\,\mathrm{mm}$ with resolution 2048 × 1944px and pixel pitch $13.5\,\mathrm{\mu m}$, the magnification is 0.0875. Only computing the aperture number with $\lambda=625\,\mathrm{nm}$ that is the highest quantum efficient point of the camera used in the Section \ref{sec:camera}, the aperture number 13 is as DOF=8cm. So the F-stop is set at 16 due to the discrete number. The particle image $\tau_p$ is 1.97. To avoid to interfere the flow, the lens and the jet nozzle are in the same plane parallel to the front plane of the investigated volume, thus the distance from the object geometrical center to the lens center $d_o$ is $35\,\mathrm{cm}$ with the camera angle at 30\textdegree. So the focal length $f$ is $28.16\,\mathrm{mm}$. The $35\,\mathrm{mm}$ lens is adapted due to discrete product models.
\subsection{Light budget computation}
The emission spectra of  a white light LED consists of a wide wavelength band from $440-760\,\mathrm{nm}$. In order to simplify the estimation of the image signal, we consider the optical power as if it is emitted at the single wavelength $\lambda=625\,\mathrm{nm}$, i.e. the following computation is based on $\lambda=625\,\mathrm{nm}$ and corresponding photon energy of $3.2\times 10^{-19}\,\mathrm{Joules}$. At this wavelength the Mie parameters $x_M$ equals 75.4, which is in the range of geometrical optics approximate computation. At the scattering angle of 90\textdegree, the light brightness at the target is $10^6 \,\mathrm{lux}$ per LED. The averaged Mie scattering coefficient is about $8.6\times 10^{-12}\,\mathrm{m}^2$. The received light energy of the CMOS detector is $2.95\times 10^{-18}\,\mathrm{Joules}$. The quantum efficiency of the CMOS detector of the camera at this photon energy is 0.95.\\
Plugging these parameters into the aforementioned model for estimating, the signal from one particle gives 8.8 counts for one LED set. If a LED can be pivoted to the cameras for acquiring a better scattering angle like , for example, 75\textdegree, the received light energy would be $1.39\times 10^{-17}\,\mathrm{Joules}$, and the particle signal on the sensor is about 41.6 counts for each LED set. The noise level of the CMOS detector is $\sigma_n=7.2$, the signal level is 10.8 and the required light density $I_c$ is 148 counts with assuming $N_{ppp}=0.01$. So the number of required LED sets is 3.5. Because all the computation are conserved, we choose to use 3 LEDs in the experiment.
\subsection{Computation result}
To summarize all the computation in this section, the steps and results are shown in the following table \ref{tab:result}.
\begin{table}
\caption{The computation results}
\begin{tabular}{cllll}
\hline\noalign{\smallskip}
Steps in & Designed & Utilizing parameters & Result & Unit  \\
Sec.\ref{sec:computationprocess} & Variables & from given condition &  &  \\
\noalign{\smallskip}\hline\noalign{\smallskip}
1 & volume & & 30×30×8 & cm\\
  & $\ell$ & & 2.2 & mm \\
  & $\tau_f$ & & 1 & ms \\
  & $U$ & & 22.65 & m/s \\
2 & $f_c$ & & 2 & kHz \\
3 & $d_p$ & & 15 & $\mathrm{\mu m}$ \\
  & $\rho_p$ & & 20-70 & $kg\cdot m^{-3}$ \\
  & $\varepsilon$ & & 0.26-0.47 & - \\
  & $\vert H_p(\varepsilon)\vert^2$ & & 0.7088 & - \\
4 & $M$ & & 0.0875 & - \\
  & $d_o$ & & 35 & cm \\
5 & $f_s$ & & 16 & - \\
  & $\tau_p$ & & 2 & pixel \\
  & $f$ & & 35 & mm \\
6 & $x_m$ & & 75.4 & - \\
7 & $\vartheta$ & & 75 & \textdegree \\
8 & $I^{(1)}_0$ & & 41.6 & counts \\
9 & $\sigma_A$ & & 10.8 & - \\
10 & $I_0$ & & 148 & - \\
  & $I_0/I^{(1)}_0$ & & 3.5 & -\\
\noalign{\smallskip}\hline
\end{tabular}
\label{tab:result}
\end{table}
Every step is computed base on the Section \ref{sec:computationprocess}. As described from Section \ref{sec:computation}, all the variables are
\section{Experiment and results}
\label{sec:results}
Experiments are conducted in a jet flow facility of the Turbulence Research Laboratory (TRL) of the Department of Mechanical Engineering at Technical University of Denmark. The TRL consists of 2 test sections -- the High resolution section and the Full Field section \cite{Ribergard2021}. All the experiments in this paper are implemented in the Full Field section.
\begin{figure*}
  \includegraphics[width=0.95\textwidth]{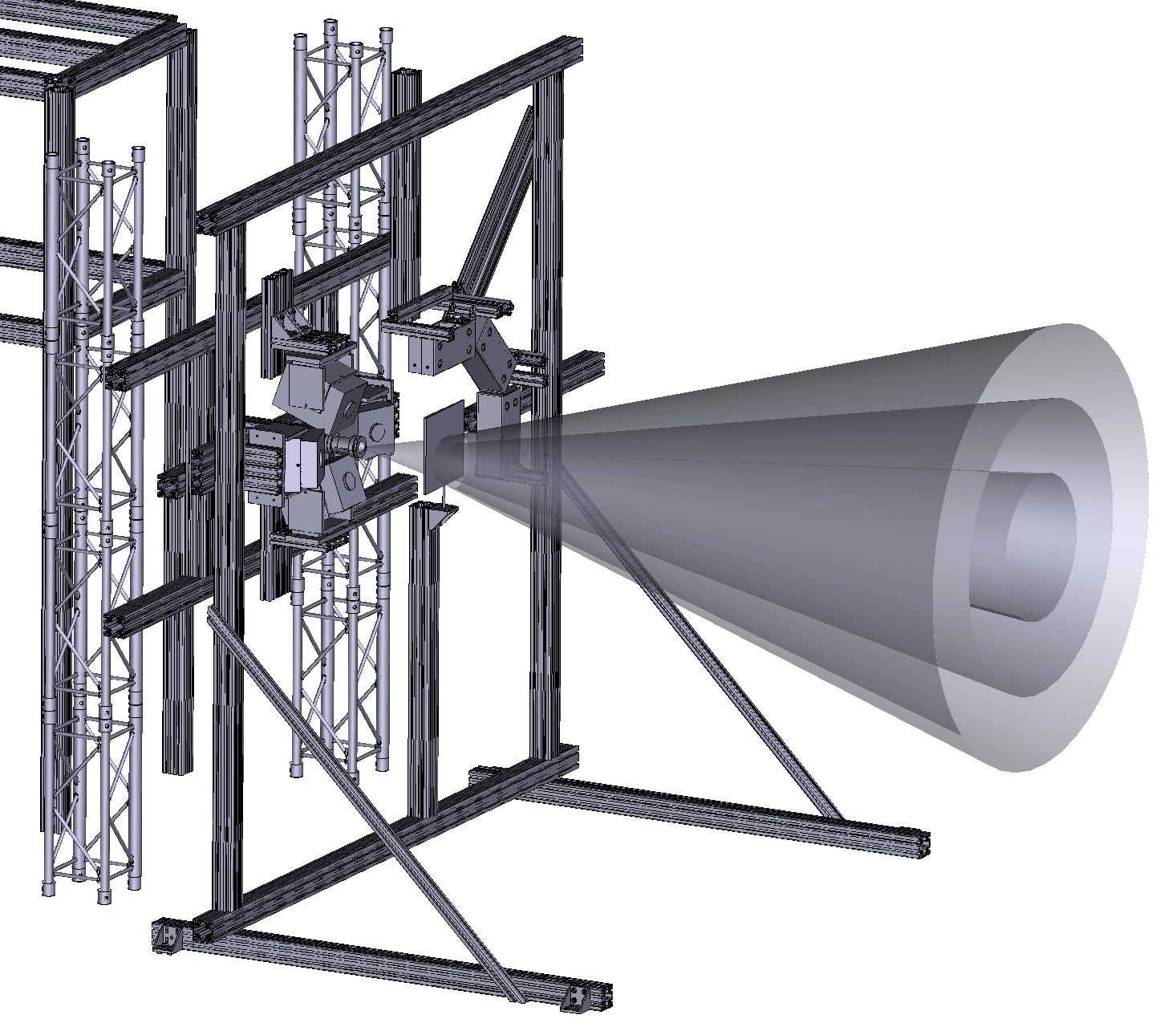}
\caption{The Full Field section in Turbulence Research Laboratory at Technical University of Denmark}
\label{fig:setup}       
\end{figure*}
\subsection{Experimental setup}
\subsubsection{Jet}
 In the Full Field test section, a jet with $10\,\mathrm{mm}$ diameter nozzle was placed in the center of injection wall. At the end of jet downstream direction, a sucking plate was mounted on the sucking wall. The whole Full Field Test Section have volume of $4480\,\mathrm{mm}$(streamwise length) ×$4110\,\mathrm{mm}$(width) ×$4930\,\mathrm{mm}$(height). Outside of the test section, there is an air recirculation controlling the air flow and the seeding density without influencing flow in the test cell.
 \begin{figure}
  \includegraphics[width=0.95\textwidth]{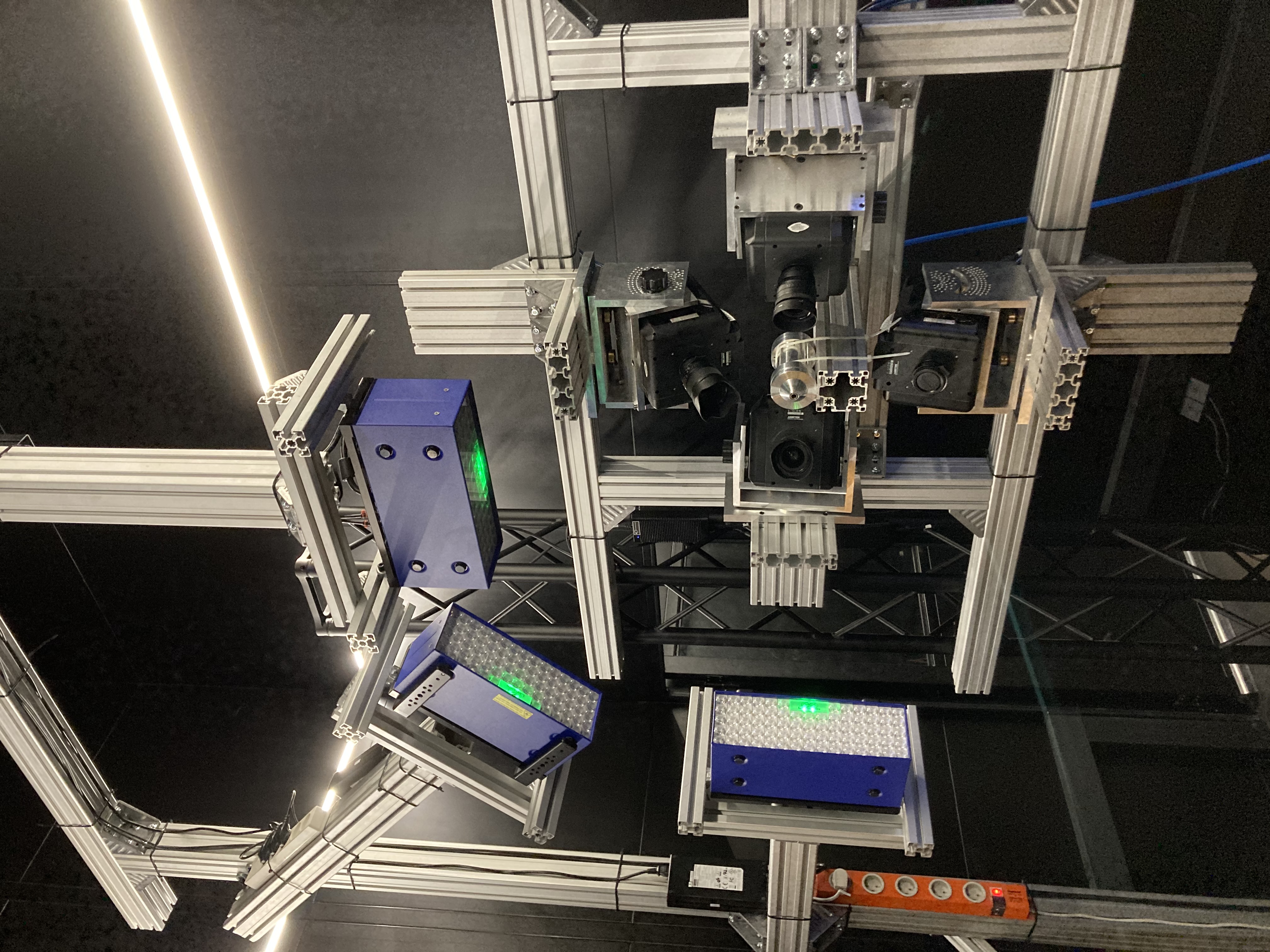}
\caption{The setup for these two measurements.}
\label{fig:setup}       
\end{figure}
\subsubsection{Cameras}
\label{sec:camera}
The visualization system that we used is a commercial LaVision PIV/PTV system, which consists of four 2048 × 1944px CMOS high-speed video cameras (v2640), a Programmable Timing Unit(PTU) and system software Davis10. The sensor size is $27.6mm\times 26.3mm$, and the pixel pitch is $13.5\mu m$ with 12-bit depth. The cameras were arranged symmetrically facing the volume of interest(VOI), as shown in Figure \ref{fig:setup}.\\
The cameras were angled at $\alpha= 30^{\circ}$ to the jet center line. The cameras were mounted on a 4 degree of freedom frame, which help to adjust the view angle at the minimum point independently. According to computation from , the lens focal length is 35mm, and the aperture is set at f\# 8.
\subsubsection{Lights}
The light sources we use in the experiment are an array of Flashlight 300 LEDs, which is produced by LaVision GmbH. These high power LEDs consist of 72 diodes each. The opening angle is 10\textdegree. In the pulsed-overdrive mode, the LEDs are operated above the nominal LED current to generate short pulses at about 5 times high light intensities than the free trigger mode. but to protect the LEDs at such high current, the duty cycle is limited to a maximum of 10\%. The brightness can be measured above $10^6\,\mathrm{lux}$ at 1m distance.\\
In order to acquire better light single, forward scattering is applied by rotating the LEDs to the camera. The forward LED angles for each camera are measured and demonstrated in the Table \ref{tab:forwardangle}\\
\begin{table}
\caption{The angle from LEDs to cameras} 
\begin{tabular}{llll}
\hline\noalign{\smallskip}
Camera & LED1 & LED2 & LED3  \\
\noalign{\smallskip}\hline\noalign{\smallskip}
Camera no.1 & 120 & 120 & 120 \\
Camera no.2 & 130 & 130 & 130 \\
Camera no.3 & 140 & 140 & 140 \\
Camera no.4 & 150 & 150 & 150 \\
\noalign{\smallskip}\hline
\end{tabular}
\label{tab:forwardangle}       
\end{table}
\subsection{Seeding}
The air-filled soap bubble generator used in this paper was developed by TSI and had been applied for several wind tunnel experiements. It consists of a $57\,\mathrm{L}$ reservoir filled with a 5\% surfactant-water solution. The surfactant was common non-color  and non-perfume pure dish-washing detergent. At bottom of the reservoir, a piston pump is used to draw and pressurize the solution through a high pressure tube with a set of 10 nozzles. The outlet diameter of the nozzles is 0.2mm, and could be blocked with special blank nozzle heads to reduce the seeding generating rate.\\
The air-filled-soap-bubble tracers have mean diameter of $14.7\,\mathrm{\mu m}$, response time  $40\,\mathrm{\mu s}$ with a dispersion of  $30\,\mathrm{\mu m}$. The bubbles can be generated at high rates of 50 tracers per $\mathrm{cm}^3$. The Stokes number of the micro-bubbles was 0.04, which could show the flow trajectory faithfully\cite{Voth2017}.
\subsubsection{Shadowgraph results of $15\,\mathrm{\mu m}$ bubbles}
\subsubsection{Phase Doppler Anemometry results of $15\,\mathrm{\mu m}$ bubbles}
\subsection{Experiment process}
The demonstrated volumetric measurements of 4D PTV are implemented into two tests in a $10\,\mathrm{mm}$ round jet. The first test is carried out at $Re=2\times 10^4$ within a domain from 30D to 38D $2930\,\mathrm{cm}^3$, which lies between the traditional volumetric domains investigated using either DEHS droplets or HSFB bubbles. The second test Reynolds number is 25,000, and the domain starts from 70D to 78D, which equals to the same volumetric domain that are presented in.\\
The data analysis is performed with LaVision software Davis 10.3. For post-processing, the images were processed using the following steps to reduce noise: 1) apply a 3 × 3 pixel Gaussian filter to remove the noise; 2) subtract a local median ( 9 × 9 pixels) to remove the background; 3) subtract a constant intensity of 10 and multiply by a constant value of 5 to increase the contrast; 4) apply a 7 × 7 pixel Gaussian filter to facilitate the particle identification process.
\subsection{Experimental Results}
\subsubsection{Light signal result}
\begin{figure*}
  \includegraphics[width=0.95\textwidth]{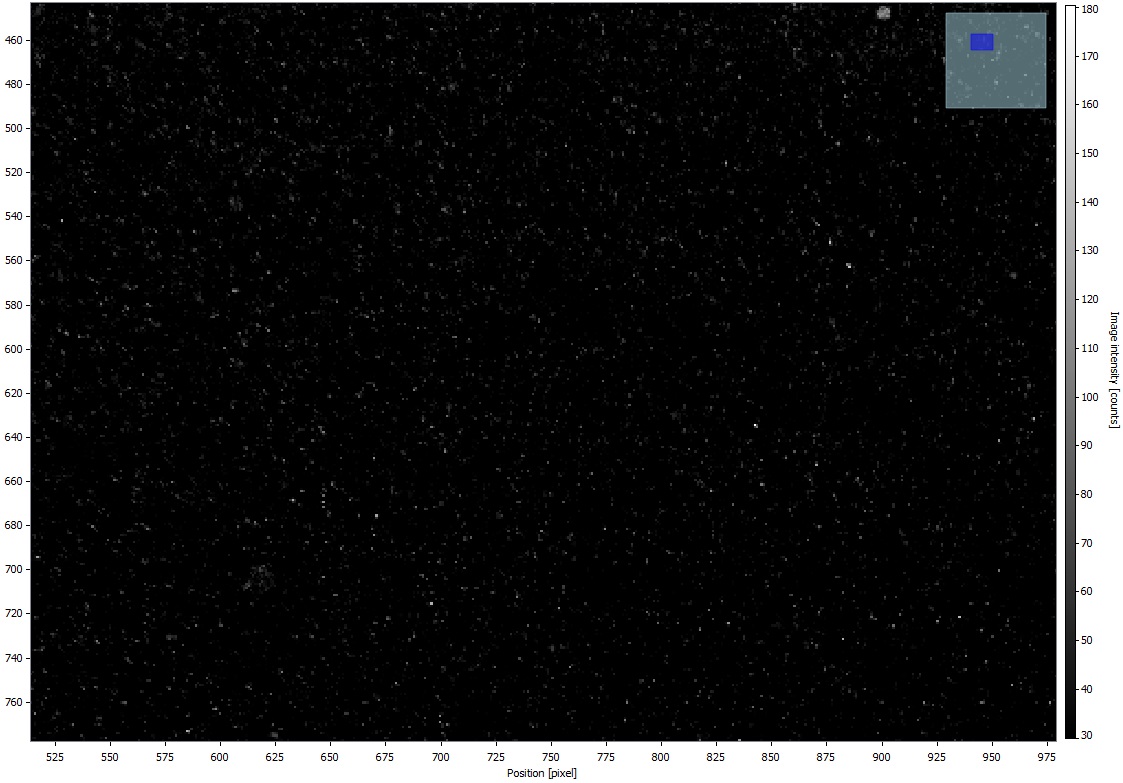}
\caption{The sub-image of the measurement.}
\label{fig:result1}       
\end{figure*}

For each snapshot, the number of the traced particles was in the range $7\times 10^4 \sim 9\times 10^4$. The signal-to-noise ratio (SNR) of the raw images is 1.6, which is higher than the estimated 1.5. This may be due to the inhomogeneous plane light that increases forward scattering, and high density of bubbles that may also contribute to the scattering light intensity \cite{Scharnowski2016}.\\
A sample image acquired by one camera was shown in Fig. \ref{fig:result1}, in which the particles can be clearly identified by the LaVision Davis system. The averaged size of particle image is 4 pixels in agreement with computation, and the signal from the camera sensor is about 80-160 counts, which is in the same level with the estimated 147 counts in section \ref{sec:computation}. The main difference may be due to the different angle of LED to different cameras.

\begin{figure*}
  \includegraphics[width=0.95\textwidth]{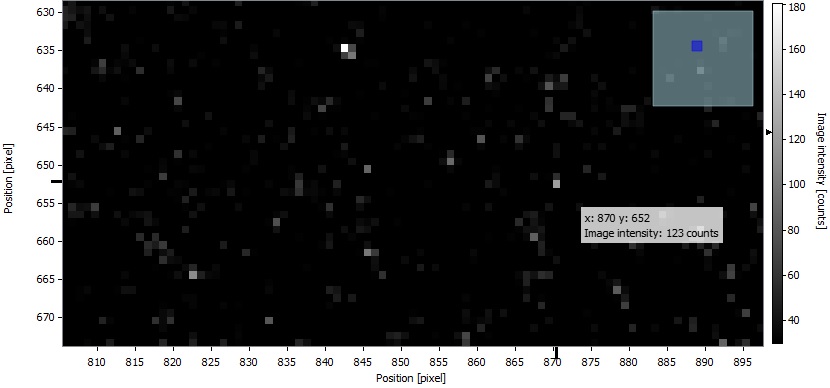}
\caption{The signal intensity is 80-160 counts.}
\label{fig:result2}       
\end{figure*}
\subsubsection{Reconstructed result}
\subsubsection{Global structure and Taylor micro-scale structure}
\section{Conclusion and discussion}
To cover the whole range from the global eddies to the smallest interested eddies, the relations of particle size, imaging, light intensity are  formulated. From the characteristic timescale and length scale of the smallest interested eddies, the tracer type can be chosen by the diameter and density. Then the light intensity can be estimated from the particle and the necessary minimum distance of the setup. With all the computed parameters, one setup for jet flow measurement was prepared for the volumetric velocimetry in the volume of $\geq 2\times10^3\,\mathrm{cm}^3$ and $\geq 9\times10^3\,\mathrm{cm}^3$. The selected seeding tracer is air-filled soap bubbles with diameter of approximately $15\,\mathrm{\mu m}$, of which the size of seeding tracer is between millimeter scale HFSB and $1\,\mathrm{\mu m}$ scale DEHS. The targeted measurement volume dimension is equivalent to the volume of HFSB, which will give a higher resolution of turbulence study. \\
The evaluation method in Section \ref{sec:method} can successfully predict the visibility of the small tracers. The Taylor micro-scale can be observed in the same time to catch the global eddies at $Re=2\times 10^4$ in the $10\,\mathrm{mm}$ jet flow. A higher Reynolds number is possible to be achieved with well designed illumination source. Also a full field volumetric velocimetry of a jet flow from 30D to 100D can be implemented with more cameras due to the limited DOF of each lens. Further developments aim to reach higher Reynolds number and larger DOF.\\
Before the experiment designs, it is better to define an index of light cost per signal counts that could help to evaluate the illumination options under the limited budget. In a larger volumes, the low cost LED lights, which pave for the basic light intensity to the threshold with the problem of illuminating unnecessary volume, can be adapted to accompany the high cost high-power laser with mirrors reflection, which increase the critical signals within the limited volume. Because the LED lights are highly limited by the distance due to the open angle , and the laser are highly limited by the reflection times due to energy loss in each reflection.\\
Our further work is to observe Taylor micro-scale in the same time to catch the global eddies at $Re=2 \times 10^4$ in the 1 cm jet flow. A higher Reynolds number is possible to be achieved with well designed illumination source.
\section{Acknowledgements}
Yisheng Zhang, Simon L. Ribergård and Clara M. Velte acknowledge financial support from the European Research council: This project has received funding from the European Research Council (ERC) under the European Unions Horizon 2020 research and innovation program (grant agreement No 803419).\\
Haim Abitan acknowledges financial support from the Poul Due Jensen Foundation: Financial support from the Poul Due Jensen Foundation (Grundfos Foundation) for this research is gratefully acknowledged.\\
We all also acknowledge Knud Erik Meyer and Niels S. Jensen for his help in our experimental implement.\\
The Department of Mechanical Engineering at the Technical University of Denmark is also acknowledged for their generous additional support in establishing the laboratory.

%
%



\end{document}